\documentclass[preprint,12pt, a4paper]{elsarticle}

\usepackage{amsmath,amsbsy}

\usepackage{subfig}
\usepackage{hyperref}

\biboptions{round, authoryear}

\usepackage{adjustbox} 

\usepackage{float}

\usepackage{threeparttable}

\usepackage{multirow}

\begin{document}

	\begin{frontmatter}
	
		\title{Comparing Analytical and Numerical Approaches to Meteoroid Orbit Determination using Hayabusa Telemetry}
        
		\author[1]{Trent Jansen-Sturgeon}
		\author[1]{Eleanor K. Sansom}
		\author[1]{Phil A. Bland}
        
		\address[1]{School of Earth and Planetary Sciences, Curtin University, GPO Box U1987, Perth, WA, 6845, Australia}

		\begin{abstract}
            Fireball networks establish the trajectories of meteoritic material passing through Earth’s atmosphere, from which they can derive pre-entry orbits. Triangulated atmospheric trajectory data requires different orbit determination methods to those applied to observational data beyond the Earth’s sphere-of-influence, such as telescopic observations of asteroids. Currently, the vast majority of fireball networks determine and publish orbital data using an analytical approach, with little flexibility to include orbital perturbations. Here we present a novel numerical technique for determining meteoroid orbits from fireball network data and compare it to previously established methods. The re-entry of the Hayabusa spacecraft, with its known pre-Earth orbit, provides a unique opportunity to perform this comparison as it was observed by fireball network cameras. 
            
            As initial sightings of the Hayabusa spacecraft and capsule were made at different altitudes, we are able to quantify the atmosphere’s influence on the determined pre-Earth orbit. Considering these trajectories independently, we found the orbits determined by the novel numerical approach to align closer to JAXA’s telemetry in both cases. Using simulations, we determine the atmospheric perturbation to become significant at $\sim$90 km; higher than the first observations of typical meteorite dropping events. 
            
            Using further simulations, we find the most substantial differences between techniques to occur at both low entry velocities and Moon passing trajectories. These regions of comparative divergence demonstrate the need for perturbation inclusion within the chosen orbit determination algorithm.
		\end{abstract}
	\end{frontmatter}

\pagebreak
\section{Introduction}
Fireball networks track meteoritic material as it transits our atmosphere. Triangulated observations of fireballs provide precise trajectories for these objects. By propagating such trajectories back in time, we can acquire orbital data for meteoroids, be it of cometary or asteroidal origin. For objects $<$10\,m diameter - typically below the resolution of telescope observations - fireball networks are currently the only method capable of delivering bulk orbital datasets for this class of solar system material. Fireball networks have an additional value in providing trajectory data that can facilitate the physical recovery of meteorites with orbits.

As of early 2018, only a mere 32 meteorites have been recovered where their observed atmospheric entry data allows an orbital trajectory to be determined with varying degrees of reliability and precision \citep{granvik_identification_2018}. The accurate knowledge of the origins of this material is vital to our understanding of Solar System formation. Differences in orbital characteristics, however slight, will be amplified with time as material is propagated back perhaps thousands, if not millions of years in order to find a match to a potential parent body or source region. Using probabilistic orbital evolution modelling techniques \citep{bottke_debiased_2002}, one can trace back a meteoroid’s determined pre-Earth orbit and probabilistically link the observed space rock to particular Near Earth Object (NEO) source regions. The mechanism triggering the migration of an object’s stable orbit, such as an unstable mean-motion orbital resonance or a close encounter with a planetary body, can be probabilistically identified. Understanding a meteoroid’s origin, and thereby uncovering a piece of recent dynamical history of the solar system, requires both accuracy and precision in the meteoroid’s initial orbit determination techniques.

One such analytical technique is outlined in Section~11 of the work by Ceplecha (1987), hereafter referred to as “C-87”. It includes two corrections to the initial velocity vector based on simplifying assumptions to determine the meteoroid’s pre-Earth orbit. An alternative approach would be a numerical propagation method - an integration-based approach that iteratively propagates a meteoroid’s initial state vector, through the most significant perturbations, back in time until the Earth’s influence is considered negligible, at which point the pre-Earth orbit is produced.

Historically, C-87 has long been used as the method of choice due to its computational ease and convenience. However, as computational power has increased, so has the viability of the numerical approach. There are at least 9 groups that publish orbital data from meteor and fireball observations, and C-87 is used by all but one of them [C-87: \citet{brown_development_2010}; \citet{colas_french_2015}; \citet{cooke_status_2012}; \citet{gural_california_2011}; \citet{madiedo_multi-station_2008}; \citet{rudawska_new_2014}; \citet{spurny_automation_2007}; \citet{wisniewski_current_2017}, Numerical: \citet{dmitriev_orbit_2015}]. The current numerical approach used by \citet{dmitriev_orbit_2015} is available as part of the standalone Meteor Toolkit package (hereafter referred to as “MT-15”) and will be compared alongside the novel numerical propagation method described in this work. This new numerical method will hereafter be referred to as “JS-19”.

Some studies have been established in the past comparing the analytical and numerical approaches to orbit determination \citet{clark_numerical_2011}, however these comparisons were conducted using published meteor observations with no pre-Earth sightings. To compare the various orbit determination methods, a real world example with well recorded data both before and after it encounters Earth’s perturbing influence, namely the pre-Earth orbit and the triangulated atmospheric trajectory respectively, would be invaluable. 

In November 2005, Japan Aerospace Exploration Agency’s (JAXA’s) Hayabusa mission successfully retrieved samples from the near-Earth asteroid 25143 Itokawa \citep{nakamura_itokawa_2011}. On its scheduled return to Earth, the Hayabusa spacecraft made several trajectory correction manoeuvres, the last being about three days before predicted re-entry over the Woomera Prohibited Area (WPA), South Australia. Following this last correction burn, the orbit was calculated using precise positional telemetry by the Deep Space Network team at NASA’s Jet Propulsion Laboratory \citep{cassell_hayabusa_2011}. On 13 June, 2010, 13:52 UT, the Hayabusa spacecraft and its return capsule made a coordinated ballistic re-entry over WPA. This re-entry was recorded by two temporary stations set up by JAXA’s ground observation team \citep{fujita_overview_2011}, four autonomous observatories of Australia’s Desert Fireball Network (DFN) \citep{borovicka_photographic_2011}, and one optical imaging station within NASA’s DC-8 airborne laboratory \citep{cassell_hayabusa_2011}. Although it is not strictly a meteoroid, the Hayabusa mission is a fitting candidate for orbit determination analysis. Its re-entry mimicked real meteoroid entry phenomena in its ballistic nature and was observed in a similar fashion to fireballs, while also possessing a ‘ground truth’ orbit from DSN telemetry.

\section{Methods}\label{sec:methods}
All orbit determination methods studied in this paper utilise the same triangulated observation data and all return identical outputs, providing an excellent setting for comparison and analysis. 

The inputs are simply the meteoroid’s initial ‘state’ taken at the highest reliable altitude that was observed. This initial state includes the absolute UTC time of observation (epoch time), and the triangulated position and velocity vectors at this time, expressed in Earth Centred Inertial (ECI) coordinates. 

The outputs are the six Classical Orbital Elements (COE’s) that describe the original orbit of the meteoroid before the gravitational influence of the Earth/Moon system at the initial observed time, or epoch time. These orbital elements are the semi-major axis ($a$), eccentricity ($e$), inclination ($i$), argument of periapsis ($\omega$), longitude of ascending node ($\Omega$), and the true anomaly ($\theta$). However, the true anomaly is generally not quoted for entry orbits if the epoch time is provided.

In this method section, we will first review the method C-87 outlined in \citet{ceplecha_geometric_1987} by presenting the approach in a more conceptual and modern setting, before going on to describe our new numerical method (JS-15). A detailed description of the Meteor Toolkit (MT-15) approach is given by \citet{dmitriev_orbit_2015}.

\subsection{Analytical Method of Ceplecha (C-87)}
\begin{figure} 
    \centering
    \includegraphics[width=\textwidth]{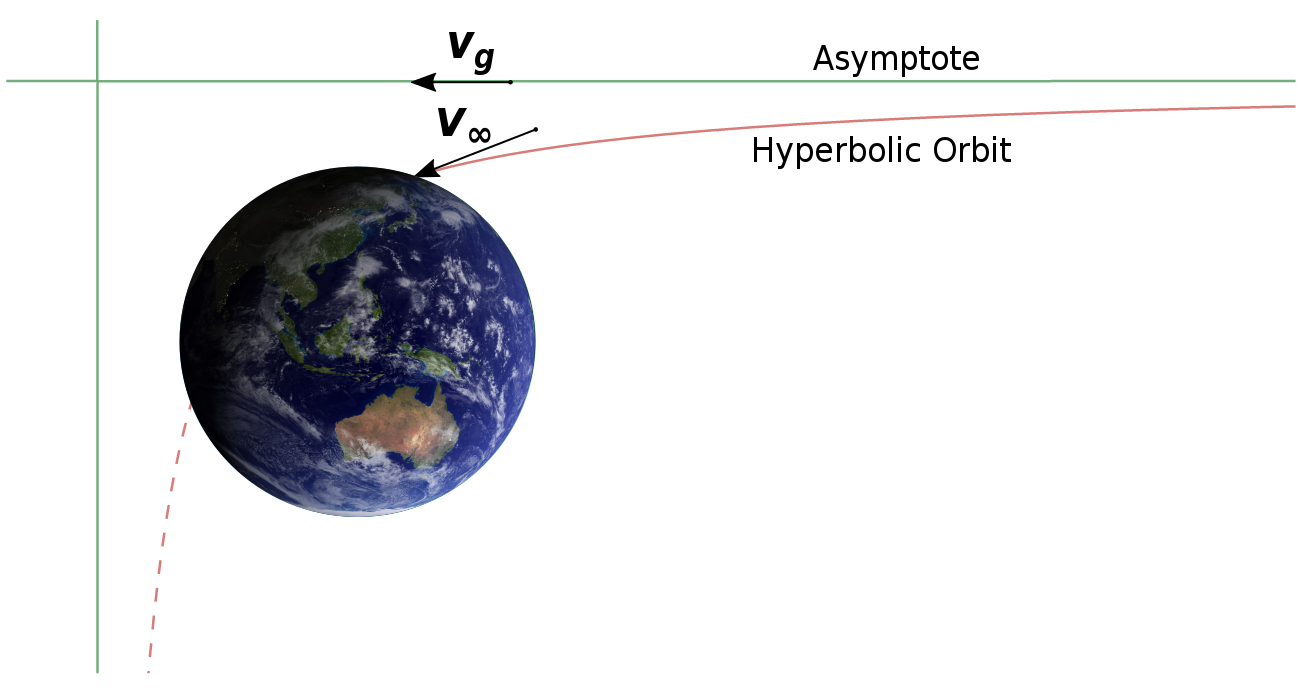}
    \caption{Ceplecha's orbital assumption in the local region of Earth, where $\vec{v}_g$ is the uninfluenced pre-Earth velocity vector and $\vec{v}_{\infty}$ is the (Earth affected) velocity vector determined outside the atmosphere using the method described in the appendix of \citet{pecina_new_1983, pecina_importance_1984}. [Globe image credit: earthobservatory.nasa.gov]}
    \label{fig:cep_concept}
\end{figure}

As first outlined in \citet{ceplecha_geometric_1987}, C-87 is based on the assumption of an initial hyperbolic collision orbit with Earth. Using the mathematical theory of conics, the hyperbolic entry orbit’s asymptote can be determined, which is taken to be the local path of the meteoroid around the Sun before Earth’s gravitational influence, as shown in Fig.~\ref{fig:cep_concept}. There are two adjustments made to the initial velocity vector that best estimate this local path relative to Earth. These adjustments are made to the magnitude and zenith angle of the initial velocity vector.

The magnitude adjustment to the initial velocity vector is two-fold; firstly to account for the atmospheric influence, and secondly to account for Earth’s gravitational attraction component. The pre-atmospheric velocity, $\vec{v}_{\infty}$, can be determined using methods described in the appendix of \citet{pecina_new_1983, pecina_importance_1984}. Using this inertial pre-atmospheric velocity, $\vec{v}_{\infty}$, and the escape velocity at that particular height, $v_{esc}$, the magnitude of the resulting geocentric velocity vector, $v_g$, can be determined as follows:

\begin{equation}\label{eq:ceplecha_correction1}
    v_g = \sqrt{\big\|\vec{v}_{\infty}\big\|^2 - v_{esc}^2} \quad \textrm{where} \quad v_{esc} = \sqrt{\frac{2\mu_e}{\big\|\vec{x}_0\big\|}}
\end{equation}

where $\mu_e = G \times m_e = 3.986005 \times 10^{14} m^3s^{-2}$ \citep{moritz_geodetic_2000} is Earth’s standard gravitational parameter, and $\vec{x}_0$ is the inertial position corresponding to the highest triangulated point.

The direction of the geocentric velocity vector is simply the direction of $\vec{v}_{\infty}$ with an adjustment to its zenith angle, $z_c$, as follows:

\begin{equation}\label{eq:ceplecha_correction2}
    a_g = a_c,\ z_g = z_c + dz_c \quad \textrm{where} \quad dz_c = 2 \arctan\left(\frac{\big\|\vec{v}_{\infty}\big\| - v_g}{\big\|\vec{v}_{\infty}\big\| + v_g} \tan\left(\frac{z_c}{2}\right)\right)
\end{equation}

where $a_c$ and $z_c$ are the local azimuth and zenith angles of the observed radiant, corrected for Earth’s rotation, and $a_g$ and $z_g$ are the azimuth and zenith angles of the geocentric radiant.

The geocentric velocity vector can now be determined from the velocity’s magnitude, azimuth and zenith angles formulated above. The resulting orbit is then calculated by transforming the geocentric position and velocity vectors, $\vec{x}_0$ and $\vec{v}_g$, into heliocentric inertial coordinates (J2000) followed by heliocentric COE’s. Notice there are no modifications to the position of the meteoroid due to Earth’s influence, or any other perturbing body, as it is assumed that any adjustment would make near negligible difference to the resulting orbital elements.

We must note that C-87 cannot determine the orbit of an entry object that had been gravitationally bound to Earth due to its primary assumption of an initial hyperbolic collision orbit with Earth. We must also note that the determination of the pre-atmospheric velocity, $\vec{v}_{\infty}$, as outlined in the appendix of \citet{pecina_new_1983}, is not as well known amongst the meteor modelling community, and is frequently mistaken as the velocity at the first triangulated point, $\vec{v}_0$ \citep{vida_modelling_2018}. This will lead to a misuse of Eq.~\ref{eq:ceplecha_correction1} and Eq.~\ref{eq:ceplecha_correction2} in calculating $\vec{v}_g$ as the Earth’s atmospheric influence will not be accounted for. In order to objectively assess the effect of omitting the pre-atmospheric velocity determination on orbital calculations, we will additionally analyse C-87 setting $\vec{v}_{\infty}$ to $\vec{v}_0$.

\subsection{Novel Numerical Method (JS-19)}
Unlike C-87, JS-19 makes no assumptions about the origin of the meteoroid and can accommodate perturbations with ease. This method effectively rewinds the clock by propagating the meteoroid’s state back in time to a point well outside the Earth’s sphere of influence. 

Modified Equinoctial Orbital Elements (EOE’s) are used to describe the meteoroid’s state as these elements avoid the singularities inherent in the COE parameterisation at zero- and ninety-degree inclinations and zero eccentricity \citep{betts_very_2000, cefola_equinoctial_1972}. The initial conditions, namely the highest reliable inertial position, $\vec{x}_{\infty}$, and velocity, $\vec{v}_{\infty}$, are converted from inertial vector coordinates into COE’s and then from COE’s into EOE’s, as outlined in Section~3.5 and Section~3.4 respectively of \citet{colasurdo_astrodynamics_2006}. These EOE’s are vectorised following the European Space Agency’s notation \citep{walker_set_1985} as:

\begin{equation}\label{eq:numerical_first}
    \vec{y} = \begin{bmatrix} p & f & g & h & k & L \end{bmatrix} ^T
\end{equation}

In order to propagate the meteoroids state elements back to its originating orbit, a dynamic model (or a set of ordinary differential equations) is needed, namely the variation of parameters on the equinoctial element model \citep{betts_very_2000}:

\begin{equation}
    \Dot{\vec{y}} = \vec{A} \ \vec{u}_{tot} + \vec{b}
\end{equation}

\begin{equation}
    \vec{A} = \begin{bmatrix} \Dot{p} \\ \Dot{f} \\ \Dot{g} \\ \Dot{h} \\ \Dot{k} \\ \Dot{L} \end{bmatrix}
    = \frac{1}{w}\sqrt{\frac{p}{\mu_e}}
    \begin{bmatrix} 0        &  2 p              &  0 \\
                    w\sin(L) &  (w+1)\cos(L) + f &  -g\beta \\
                    w\cos(L) &  (w+1)\sin(L) + g &  -f\beta \\
                    0        &  0                &  s^2\cos(L)/2 \\
                    0        &  0                &  s^2\cos(L)/2 \\
                    0        &  0                &  \beta \end{bmatrix}
\end{equation}

\begin{equation}
    \vec{b} = \begin{bmatrix} 0 & 0 & 0 & 0 & 0 & \sqrt{\mu_ep}(w/p)^2 \end{bmatrix} ^T
\end{equation}

where $\vec{A}$ is the state rate matrix, $\vec{b}$ is the state rate constant, and $\vec{u}_{tot}$ is the total perturbing acceleration in the body frame [radial, tangential, normal]. Also $w$, $s$, $r$, and $\beta$ are some shorthand notations of common expressions:

\begin{equation}
\begin{split}
    w = 1& + f\cos(L) + g\sin(L), \quad s^2 = 1 + h^2 + k^2 \\
    &r = p/w, \quad \beta = h\sin(L) - k\cos(L)
\end{split}
\end{equation}

For accurately determining the original orbit of incoming meteoroids, perturbations need to be added to this dynamic model. However, as there will be relatively minimal net movement of the meteoroid through time, the only perturbations that would non-negligibly affect the resulting orbit are those produced by the Earth/Moon system. These include the atmospheric drag, third body gravitational and zonal harmonic perturbations. 

The first zonal harmonic (J2) perturbation is due to the Earth’s oblate shape, and is about three times the magnitude of the next zonal harmonic \citep{moritz_geodetic_2000}. Therefore, the Earth’s J2 zonal harmonic perturbation is the only one considered, and is calculated in the body frame as follows \citep{kechichian_treatment_1997}:

\begin{equation}
    \vec{u}_{J_2} = \frac{-3\mu_eJ_2R_e^2}{r^4s^4} \begin{bmatrix} (s^4 - 12\beta^2)/2 \\ 4\beta (h\cos(L) + k\sin(L)) \\ 2\beta (2 - s^2) \end{bmatrix}
\end{equation}

where $J_2 = 1.08263 \times 10^{-3}$ \citep{moritz_geodetic_2000} is the dynamical form factor of the Earth, and $R_e = 6371.0~km$ \citep{moritz_geodetic_2000} is the Earth’s mean radius.

The Newtonian third body perturbation equation has been shown to often promote substantial numerical errors due to the significantly different magnitude of the terms involved \citep{battin_introduction_1999}. To avoid this numerical inaccuracy, the following equation \citep{betts_optimal_2003} is used to model third body perturbations in the inertial (J2000) frame:

\begin{equation}
\begin{split}
    \vec{u}_{tb} = -\mu_{tb} \frac{\vec{x}_m + f\vec{\rho}_{tb}}{\big\|\vec{x}_m - \vec{\rho}_{tb}\big\|}& \\
    \textrm{where} \quad f = \frac{3q + 3q^2 + q^3}{1 + (1 + q)^{3/2}} \quad \textrm{and} \quad q =& \vec{x}_m \cdot \frac{\vec{x}_m - 2\vec{\rho}_{tb}}{\big\|\vec{\rho}_{tb}\big\|^2}
\end{split}
\end{equation}

where $\vec{x}_m$ is the position of the meteoroid, $\vec{\rho}_{tb}$ is the position of the third body, and $\vec{u}_{tb}$ is the standard gravitational parameter of the third body. 

Finally, while the atmospheric drag acceleration is a fairly standard formula, the density of air in the upper atmosphere is not. The density in this region varies with not only height, but latitude, longitude, time, and solar activity. To incorporate all these subtle effects, we utilised the empirical NRLMSISE-00 atmospheric model \citep{picone_nrlmsise-00_2002} to calculate the atmospheric density ($\rho_{air}$) within our drag equation:

\begin{equation}\label{eq:numerical_last}
    \vec{u}_{drag} = \frac{-\rho_{air} C_d S \big\|\vec{v}_{rel}\big\| \vec{v}_{rel}}{2 M}
\end{equation}

where $M$ is the mass of the meteoroid, $C_d$ is the drag coefficient, $S$ is the meteoroid’s cross-sectional area, and $\vec{v}_{rel}$ is the meteoroid’s velocity vector relative to the surrounding atmospheric air. Note that like the third body perturbation, the atmospheric drag perturbation needs a coordinate transformation into the body frame to be used in the dynamic model. 

Now that the dynamic model is established (Eq.~\ref{eq:numerical_first}-\ref{eq:numerical_last}), a numerical integrator is needed to propagate the meteoroid’s state variables through time. We have chosen a Runge-Kutta Dormand-Prince (RKDP) \citep{dormand_family_1980} method for the integration due to its ability to constrain relative errors by internally controlling step size; a new approach to numerical fireball orbit modelling. Additionally, it supports a good accuracy to computation ratio, namely fifth order accuracy for six function evaluations per step. 

The RKDP method computes and compares a fourth and fifth order Runge-Kutta solution in parallel to determine whether the current time step is sufficiently small. If the difference between the solutions exceeds the error bounds, then the time step is decreased (by 1/10) and the RKDP is re-run on the current iteration step. If this difference is much smaller than the error bounds, the current solution is taken and the time step is increased (by 1/10) for the next RKDP iteration. The coefficients of the RKDP were chosen to minimise the error of the fifth order solution, therefore it is this solution that is used in the next step of the integration procedure.

Starting with an initial step size estimate of a tenth of a second, we use the RKDP iterative integration process to propagate the meteoroid’s Earth centred inertial EOE’s, $y_{initial}$, to the edge of the Earth’s sphere of influence (SOI), where the coordinates are converted into the Sun centred inertial frame (J2000). The integration process is then continued until the meteoroid has propagated to ten SOI, upon which the Earth/Moon perturbations are removed from the dynamic model and the meteoroid is propagated back to epoch time. The resulting orbital elements, $y_{final}$, reflect the meteoroid’s original orbit around the Sun expressed in J2000 coordinates, and can be trivially converted to COE’s as described in Section~3.4 of \citet{colasurdo_astrodynamics_2006}. 

Discontinuities can arise when switching between geocentric and heliocentric reference frames. To avoid such a discontinuity at the limit of Earth’s SOI, the Sun and Moon are considered perturbations when in the geocentric frame, while the Earth and Moon are considered perturbations within the heliocentric frame.

The JS-19 method described above is similar to that of MT-15 \citep{dmitriev_orbit_2015}, but differs in the choice of state representation, integration method, and error handling.

\section{Results and Discussion}
To properly compare these methods for their accuracy, an example object with both a measured orbit and a measured bright flight entry would be invaluable for analysis. The re-entry of the Hayabusa mission constitutes an excellent calibration event in this regard, with a measured pre-Earth rendezvous orbit, as determined by the spacecraft’s navigational systems, and an observed re-entry trajectory, as published in \citet{borovicka_photographic_2011}.

JAXA’s engineering team kindly provided their orbital telemetry data for the Hayabusa mission (through personal communication) at 2010-06-09T06:04:00.0 UTC, just after its final correction manoeuvre (TCM-4), in the form of a J2000 equatorial (Earth-Centred Inertial) state vector. This position and velocity state is easily converted into the following COE’s:

\begin{equation}
    \begin{bmatrix} R_x \\ R_y \\ R_z \\ V_x \\ V_y \\ V_z \end{bmatrix}
    = \begin{bmatrix} -1.074047355 \times 10^6\,km \\ 1.232756795 \times 10^6\,km \\ 0.935509892 \times 10^6\,km \\ 2.751442755\,km/s \\ -3.231296260\,km/s \\ -2.442756954\,km/s \end{bmatrix} 
    \xrightarrow{} 
    \begin{bmatrix} a \\ e \\ i \\ \omega \\ \Omega \\ \theta \end{bmatrix}
    = \begin{bmatrix}  1.32381\,AU \\ 0.25732 \\ 1.68383\,^{\circ} \\ 147.47773\,^{\circ} \\ 82.46569\,^{\circ} \\ 27.71211\,^{\circ} \end{bmatrix}
\end{equation}

Triangulated positions of the Hayabusa re-entry from ground based observations are detailed in \citet{borovicka_photographic_2011}. Two reduced trajectories given in this work are for the observed re-entry of the spacecraft and for the capsule; these can be used as two separate cases for orbit determination method comparisons. The tabulated triangulated positions and time in \citet{borovicka_photographic_2011} are used to determine the velocity, thereby defining the initial conditions of the luminous trajectories. In both the spacecraft and capsule cases, the numerical propagation methods will integrate the corresponding object back to the time of telemetry reading for consistent orbital comparisons. Since C-87 does not consider any perturbations, an epoch change would simply require a two body propagation, altering only the orbit’s anomaly ($\theta$). As this sixth element is not needed for orbit comparison analysis, the epoch re-calculation is not necessary.

In the two cases, we compare the orbit determined using the three different methods (C-87, MT-15, and JS-19). As the atmospheric correction to the initial velocity vector \citep{pecina_new_1983, pecina_importance_1984} used in C-87 is not always applied \citep{vida_modelling_2018}, we shall additionally provide the orbital results of C-87 by equating $\vec{v}_{\infty}$ and $\vec{v}_0$ (see Section~\ref{sec:methods}) to assess the effects.

\subsection{Hayabusa’s orbit determined from the spacecraft’s re-entry}
The initial position vector and corresponding initial time of the spacecraft can be taken directly from Table~2 of \citet{borovicka_photographic_2011} at a height of $99.88\,km$. However, as there was no given radiant vector describing the spacecraft’s velocity, the initial velocity vector of the spacecraft was deduced using a straight line least squares approach on the first three\footnote{The orbits derived by fitting different numbers of initial data points were analysed and compared with similar results. For simplicity, only one case is documented in this paper.} triangulated positional data points with timing in Table~2 of \citet{borovicka_photographic_2011}. 

Additionally, the atmospheric perturbation model requires an estimated mass and cross-sectional area of the object to more accurately model the aerodynamics. While the mass and shape of the spacecraft are relatively well documented to be $415\,kg$ and $1.5\,m \times 1.5\,m \times 1.05\,m$ cube respectively, the orientation of the spacecraft with respect to the atmosphere is more uncertain. This leaves us to assume the spacecraft’s cross-sectional area corresponds to its most aerodynamically stable orientation.

\begin{figure}
    \centering
    \includegraphics[width=\textwidth]{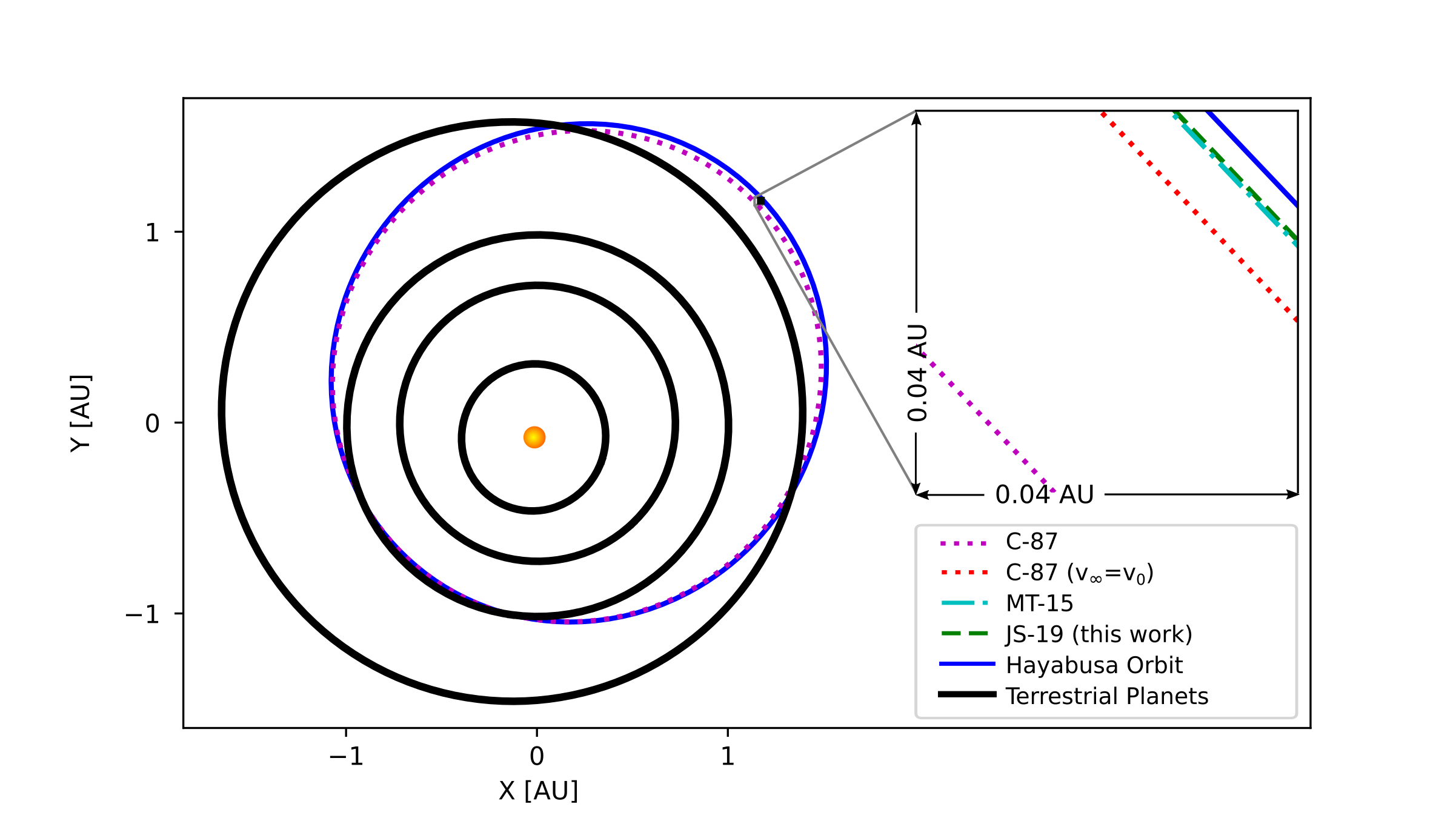}
    \caption{A comparison between Hayabusa’s heliocentric orbit (as determined from telemetry) and the spacecraft orbit calculated using Ceplecha’s analytical method (C-87), Dmitriev’s numerical method (MT-15) and the new numerical method outlined in this work (JS-19), as projected on the plane of the ecliptic. Also featuring the inner terrestrial planets as references. Included is an enlarged view around the communal aphelion to emphasise the orbital discrepancies. Note the difference in CAM orbits using different pre-atmospheric velocity assumptions (see Section~\ref{sec:methods}).}
    \label{fig:spacecraft_orbits}
\end{figure}

Using these initial conditions, the heliocentric orbit is calculated using all three methods and are compared to the orbit derived from the spacecraft’s navigation system (Table~\ref{table:spacecraft_orbits}; Fig.~\ref{fig:spacecraft_orbits}). The Southworth and Hawkins similarity criterion \citep{southworth_statistics_1963} is included in Table~\ref{table:spacecraft_orbits} as a quantitative measure of the orbital difference between JAXA’s telemetric orbit and the orbit determined using the respective methods.

\begin{table}[h!] 
    \centering
    \caption{The calculated heliocentric classical orbital elements for the Hayabusa satellite’s Earth rendezvous as compared to the telemetric orbital data at T = 2010-06-09T06:04:00.0 UTC. Note: the errors are determined assuming 10 m/s error on the initial velocity magnitude, as discussed in Section~\ref{sssec:precision}.}
    \begin{threeparttable}
    \begin{adjustbox}{width=\textwidth}
    \begin{tabular}{ |l||c|c|c|c|c| }
        \hline
        \textbf{Heliocentric}  & & & & &  \\ 
        \textbf{Orbital} & \textbf{Telemetry} & \textbf{C-87} & \textbf{C-87 ($\vec{v}_{\infty} = \vec{v}_0$)}\tnote{c} & \textbf{MT-15} & \textbf{JS-19} \\ 
        \textbf{Elements} & \textbf{Data}\tnote{b} & \citep{ceplecha_geometric_1987} & \citep{ceplecha_geometric_1987} & {\small\citep{dmitriev_orbit_2015}} & (this work) \\ 
        (ECLIPJ2000) & & & & & \\ 
        \hline \hline \multirow{2}{*}{a [km]}
        & \multirow{2}{*}{1.32381}   & 1.30395     & 1.32000     & 1.32241     & 1.32265     \\ 
        &                            & $\pm$ 0.003 & $\pm$ 0.003 & $\pm$ 0.001 & $\pm$ 0.003 \\ 
        \hline \multirow{2}{*}{e}
        & \multirow{2}{*}{0.25732}   & 0.24589     & 0.25472     & 0.25646     & 0.25654     \\ 
        &                            & $\pm$ 0.002 & $\pm$ 0.002 & $\pm$ 0.0007& $\pm$ 0.002 \\
        \hline \multirow{2}{*}{i [deg]}        
        & \multirow{2}{*}{1.68383}   & 1.64028     & 1.67009     & 1.68203     & 1.68367     \\ 
        &                            & $\pm$ 0.007 & $\pm$ 0.007 & $\pm$ 0.002 & $\pm$ 0.007 \\ 
        \hline \multirow{2}{*}{$\omega$ [deg]}
        & \multirow{2}{*}{147.47773} & 147.96599   & 147.67417   & 147.48000   & 147.52451   \\ 
        &                            & $\pm$ 0.2   & $\pm$ 0.2   & $\pm$ 0.07  & $\pm$ 0.2   \\ 
        \hline \multirow{2}{*}{$\Omega$ [deg]}
        & \multirow{2}{*}{82.46569}  & 82.34476    & 82.34414    & 82.46687    & 82.46664    \\
        &                            & $\pm$ 0.001 & $\pm$ 0.001 & $\pm$ 0.0002& $\pm$ 0.002 \\ 
        \hline Similarity     & \multirow{2}{*}{N/A} & \multirow{2}{*}{0.01178} & \multirow{2}{*}{0.00269} & \multirow{2}{*}{0.00087} & \multirow{2}{*}{0.00082} \\
        Criterion\tnote{a} & & & & &\\
        \hline
    \end{tabular}
    \end{adjustbox}
    \begin{tablenotes}\begin{tiny}
        \item[a] \citet{southworth_statistics_1963} similarity criterion as compared to the telemetry data.
        \item[b] Obtained through private communication with JAXA’s engineering team.
        \item[c] The pre-encounter velocity, $\vec{v}_{\infty}$, uses the velocity calculated at first observed point ($\vec{v}_0$). See Section~\ref{sec:methods} for details. 
    \end{tiny}\end{tablenotes}
    \end{threeparttable}
    \label{table:spacecraft_orbits}
\end{table}

\begin{figure}
    \centering
    \includegraphics[width=\textwidth]{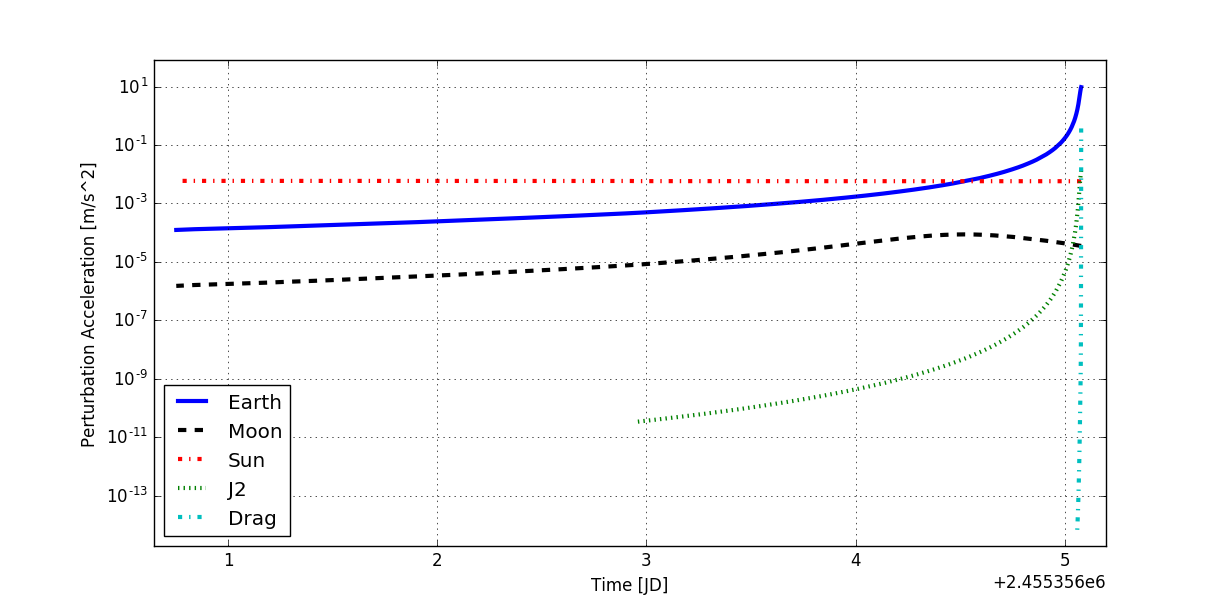}
    \caption{Selected perturbations over the Hayabusa's orbit from the final correction manoeuvre (TCM4) until Earth rendezvous. Note: the Earth’s J2 and atmospheric drag perturbations are considered negligible outside the Earth’s SOI at 924,000 km, and above the exosphere at 10,000 km respectively.}
    \label{fig:spacecraft_perturbations}
\end{figure}

The perturbations included in JS-19 are those of the Earth, Moon, and Sun’s gravity, the Earth’s first zonal harmonic ($J_2$), and the atmospheric drag. These are the non-reversible, significant perturbing effects. Their respective strengths are calculated iteratively during the backward integration of JS-19 and are represented in Fig.~\ref{fig:spacecraft_perturbations}. C-87 neglects the majority of these influences.

\subsection{Hayabusa’s orbit determined from the capsule’s re-entry}
The second interesting case is that of the Hayabusa capsule’s re-entry; it is only distinguished from the other parts of ablating spacecraft much lower down in the atmosphere ($\sim65\,km$ altitude). Although the capsule has already decelerated heavily by this point, its mass and cross-sectional diameter are very well documented to be $20\,kg$ and $40\,cm$ respectively. This sets us up for an excellent comparative study as to the effects of the atmospheric perturbation on the resulting orbital predictions.

\begin{figure}
    \centering
    \includegraphics[width=\textwidth]{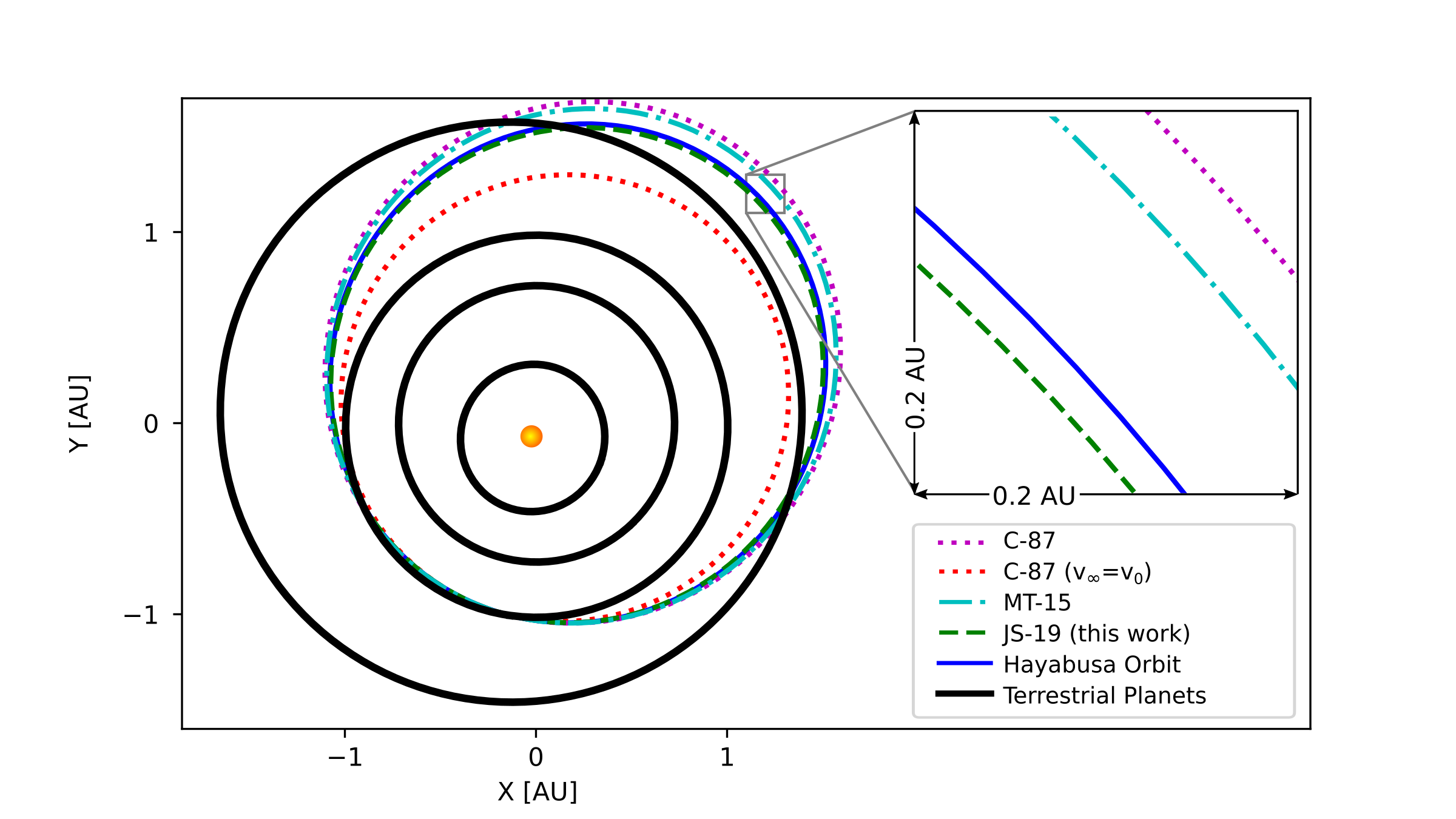}
    \caption{A comparison between Hayabusa’s orbit (as determined from telemetry) and the capsule’s orbit calculated by Ceplecha’s analytical method (C-87), Dmitriev’s numerical method (MT-15) and the new numerical method outlined in this work (JS-19), as projected on the plane of the ecliptic. Included is an enlarged view around the communal aphelion to emphasise the orbital discrepancies. Note the difference in C-87 orbits using different pre-atmospheric velocity assumptions (see Section~\ref{sec:methods}).}
    \label{fig:capsule_orbits}
\end{figure}

The initial inputs for this case originated from Table~3 of \citet{borovicka_photographic_2011}. The initial position corresponds to the highest recorded sighting of the capsule with timing, corresponding to $64.71\,km$. The initial velocity vector is deduced from the first two given position data points with timing. Note a straight line least squares fit was not attempted here as the capsule was already in a state of high deceleration. Also note that the telemetry provided for the capsule is approximated as we have not accounted for the small delta-v used in capsule ejection $3\, hours$ prior to re-entry \citep{cassell_hayabusa_2011}. Again, two cases of C-87 are assessed using different initial velocity approaches. The comparison of orbital results are shown in Table~\ref{table:capsule_orbits} 2 and Fig.~\ref{fig:capsule_orbits}.

\begin{table}[h!] 
    \centering
    \caption{The calculated heliocentric classical orbital elements for the Hayabusa capsule’s Earth rendezvous as compared to the telemetric orbital data at T = 2010-06-09T06:04:00.0 UTC. Note: the errors are determined assuming 10 m/s error on the initial velocity magnitude, as discussed in Section~\ref{sssec:precision}.}
    \begin{threeparttable}
    \begin{adjustbox}{width=\textwidth}
    \begin{tabular}{ |l||c|c|c|c|c| }
        \hline
        \textbf{Heliocentric}  & & & & &  \\ 
        \textbf{Orbital} & \textbf{Telemetry} & \textbf{C-87} & \textbf{C-87 ($\vec{v}_{\infty} = \vec{v}_0$)}\tnote{c} & \textbf{MT-15} & \textbf{JS-19} \\ 
        \textbf{Elements} & \textbf{Data}\tnote{b} & \citep{ceplecha_geometric_1987} & \citep{ceplecha_geometric_1987} & {\small\citep{dmitriev_orbit_2015}} & (this work) \\ 
        (ECLIPJ2000) & & & & & \\ 
        \hline \hline \multirow{2}{*}{a [km]}
        & \multirow{2}{*}{1.32381}   & 1.38633     & 1.17873     & 1.36699     & 1.31322     \\ 
        &                            & $\pm$ 0.003 & $\pm$ 0.003 & $\pm$ 0.001 & $\pm$ 0.003 \\ 
        \hline \multirow{2}{*}{e}
        & \multirow{2}{*}{0.25732}   & 0.28928     & 0.16954     & 0.27995     & 0.25160     \\ 
        &                            & $\pm$ 0.002 & $\pm$ 0.002 & $\pm$ 0.0007& $\pm$ 0.002 \\
        \hline \multirow{2}{*}{i [deg]}        
        & \multirow{2}{*}{1.68383}   & 1.75327     & 1.32041     & 1.73243     & 1.64657     \\ 
        &                            & $\pm$ 0.007 & $\pm$ 0.007 & $\pm$ 0.002 & $\pm$ 0.007 \\ 
        \hline \multirow{2}{*}{$\omega$ [deg]}
        & \multirow{2}{*}{147.47773} & 150.05468   & 138.57245   & 149.13093   & 146.99422   \\ 
        &                            & $\pm$ 0.2   & $\pm$ 0.2   & $\pm$ 0.06  & $\pm$ 0.2   \\ 
        \hline \multirow{2}{*}{$\Omega$ [deg]}
        & \multirow{2}{*}{82.46569}  & 82.34249    & 82.35312    & 82.44881    & 82.47087    \\
        &                            & $\pm$ 0.001 & $\pm$ 0.001 & $\pm$ 0.0002& $\pm$ 0.002 \\ 
        \hline Similarity     & \multirow{2}{*}{N/A} & \multirow{2}{*}{0.03413} & \multirow{2}{*}{0.09428} & \multirow{2}{*}{0.02394} & \multirow{2}{*}{0.00615} \\
        Criterion\tnote{a} & & & & &\\
        \hline
    \end{tabular}
    \end{adjustbox}
    \begin{tablenotes}\begin{tiny}
        \item[a] \citet{southworth_statistics_1963} similarity criterion as compared to the telemetry data.
        \item[b] Obtained through private communication with JAXA’s engineering team.
        \item[c] The pre-encounter velocity, $\vec{v}_{\infty}$, uses the velocity calculated at first observed point ($\vec{v}_0$). See Section~\ref{sec:methods} for details. 
    \end{tiny}\end{tablenotes}
    \end{threeparttable}
    \label{table:capsule_orbits}
\end{table}

The drastic difference between the predicted orbits of the Hayabusa capsule is primarily due to how the atmosphere is considered by the various methods. This orbital discrepancy really highlights the importance of a well modelled atmospheric perturbation influence in the orbit determination algorithm, especially for those objects initially observed at lower altitudes, such as some meteorite dropping fireballs.

\subsection{Atmospheric Influence}
The significant difference between the determined pre-Earth orbits of the Hayabusa capsule is due to the handling of perturbations. The most dominant perturbation in this case is the atmosphere. To assess the altitude at which the atmospheric influence on the orbit diminishes, we compared JS-19 (that accounts for the atmosphere) to C-87 where $\vec{v}_{\infty} = \vec{v}_0$ (that negates the atmosphere). The initial conditions for these comparisons were determined using JS-19; i.e. the Hayabusa capsule was integrated back along its re-entry path to a specified altitude at which point C-87 ($\vec{v}_{\infty} = \vec{v}_0$) was initiated alongside JS-19. The orbital difference between these two orbit determination methods from these initiation altitudes were then determined using Southworth and Hawkins similarity criterion \citep{southworth_statistics_1963}.

\begin{figure}
    \centering
    \includegraphics[width=\textwidth]{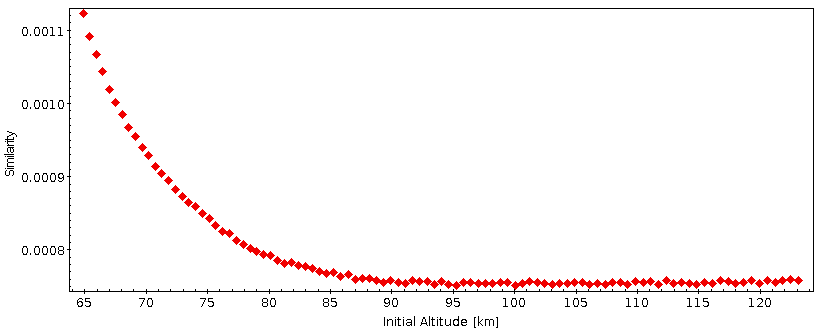}
    \caption{The orbital similarity between C-87 ($\vec{v}_{\infty} = \vec{v}_0$) and the new numerical approach at different initial altitude states, according to the Southworth and Hawkins similarity criterion \citep{southworth_statistics_1963}.}
    \label{fig:atm_influence}
\end{figure}

Fig.~\ref{fig:atm_influence} reveals a couple of interesting features about the comparative nature of the two orbit determination methods. Firstly, the similarity is shown to converge to a fixed value $\sim90\,km$ altitude, indicating the atmospheric influence on the orbit diminishes at this point. Many meteorite dropping events are not observed before this altitude, and are thus already experiencing significant atmospheric drag. However, the object’s physical characteristics, such as mass, shape, and density, would directly influence the magnitude of this atmospheric perturbation, and hence this $90\,km$ convergence altitude is specific to the case of the Hayabusa capsule. Variation to this altitude for other events requires further investigation.

Secondly, the apparent asymptote at high altitudes is non-zero. This is due to the continuing effects of the larger scale perturbations acting on the object, namely the Earth flattening and third body affects. While the magnitude of the Earth flattening perturbation drops off relatively quickly, the third body perturbations continue to influence the object’s orbit over the duration of the integration.

\subsection{Error Analysis}
For the orbital results to be validated and properly compared, their errors must be identified and quantified. These errors originate from a variety of sources, which can be factored into two groups; the observational errors and the model errors. 

The observational errors are simply the uncertainties associated with the epoch time, the initial triangulated position vector and the initial determined velocity vector before the orbital calculations begin. While the epoch time and positional errors are merely the uncertainties in the measurement data, the velocity errors are not so straightforward. The directional errors of the velocity are calculated by considering the triangulated positional radiant data as a whole, therefore minimising the potential errors in the radiant entry angle. On the other hand, the errors in velocity magnitude are determined by referring to the velocity scatter at the beginning of the object’s observable bright flight, before the atmosphere presents a significant resistive influence.

The model errors are the uncertainties introduced within the orbit determination method itself, such as the imperfect nature of the state equations in representing meteoroid flight (small perturbations missed etc.), performing discrete time integration using the Dormand-Prince integrator (bounded at 1mm per time-step), and the use of coordinate transforms\footnote{All coordinate transforms were performed using version 1.3 of Python’s Astropy module.}. Despite model uncertainties being small with respect to observational errors, their inclusion must be considered for a robust analysis. Combining all these uncertainties gives the overall error, or precision, of the results.

\subsubsection{Precision}\label{sssec:precision}
The precision of the orbit determination methods is primarily controlled by the error in the initial velocity magnitude\footnote{Using typical errors calculated by atmospheric trajectory modelling.}. The epoch time error, initial triangulated position error and the model errors combined cause an orbital uncertainty three orders of magnitude smaller than the initial velocity magnitude error alone. The initial velocity directional error is somewhat more influential on the resulting orbital errors, but still between one and two orders of magnitude smaller than the orbital uncertainty caused by the initial velocity magnitude error.

No individual position errors were provided for the triangulation results in the original paper \citep{borovicka_photographic_2011}. In order to perform a general error analysis, a velocity magnitude error of $10\,m/s$ was assumed, with error results given alongside the corresponding orbital elements in Table~\ref{table:spacecraft_orbits} and Table~\ref{table:capsule_orbits}. Other velocity magnitude errors were considered and found to scale roughly linearly to the resulting orbital errors; i.e. multiplying the velocity magnitude error by two causes the orbit uncertainty to double.

The errors on JS-19 are calculated using a Monte Carlo approach to handle the non-linearity of the included perturbations, where the error on the initial velocity magnitude can be transformed into errors on the final orbital elements. The reliability of these errors were confirmed through repeated Monte Carlo trials each consisting of one thousand particles. The error on the orbit determined by C-87 was also calculated using a Monte Carlo approach, however the error determined by MT-15 uses a series of covariance transforms throughout the algorithm. This covariance approach linearises the error at each step, and therefore does not account for any significant non-linear effects, such as a close encounter with the moon.

Table~\ref{table:spacecraft_orbits} and Table~\ref{table:capsule_orbits} reveal that the orbital precision of C-87 and JS-19 only differ significantly in their longitude of ascending node, $\Omega$. This small discrepancy is due to C-87 assuming that the meteoroid’s original (pre-perturbed) $\Omega$ is simply the Earth’s heliocentric longitude at the time of initial contact, which does not completely account for the Earth’s gravitational influence on the meteoroid’s trajectory. \citet{clark_numerical_2011} suggest that “the very tight uncertainties often reported for $\Omega$ are far too aggressive, and should be minimally expanded to incorporate this discrepancy”. This is clearly demonstrated by comparing the true $\Omega$ to the analytically and numerically determined $\Omega$ in Table~\ref{table:spacecraft_orbits} and Table~\ref{table:capsule_orbits}, highlighting the imprecise assumption that C-87 employs.

\subsubsection{Accuracy}\label{sssec:accuracy}
While the precision describes the spread of orbital results around the determined solution, the accuracy is a measure of how close that solution comes to the true orbit, or in our case, the orbit as determined using the spacecraft’s navigational systems. This error can be quantified by calculating the difference between the true orbital elements and the determined orbital elements. However, a more robust and encompassing measure of the determination method’s accuracy is by employing the similarity criterion \citep{southworth_statistics_1963}. As shown in Table~\ref{table:spacecraft_orbits} and Table~\ref{table:capsule_orbits}, the new numerical approach consistently produces more accurate orbital results. This comparison of accuracy has also been demonstrated visually in Fig.~\ref{fig:spacecraft_orbits} and Fig.~\ref{fig:capsule_orbits}.

\subsection{Relative Similarity}
\begin{figure}
    \centering
    \includegraphics[width=\textwidth]{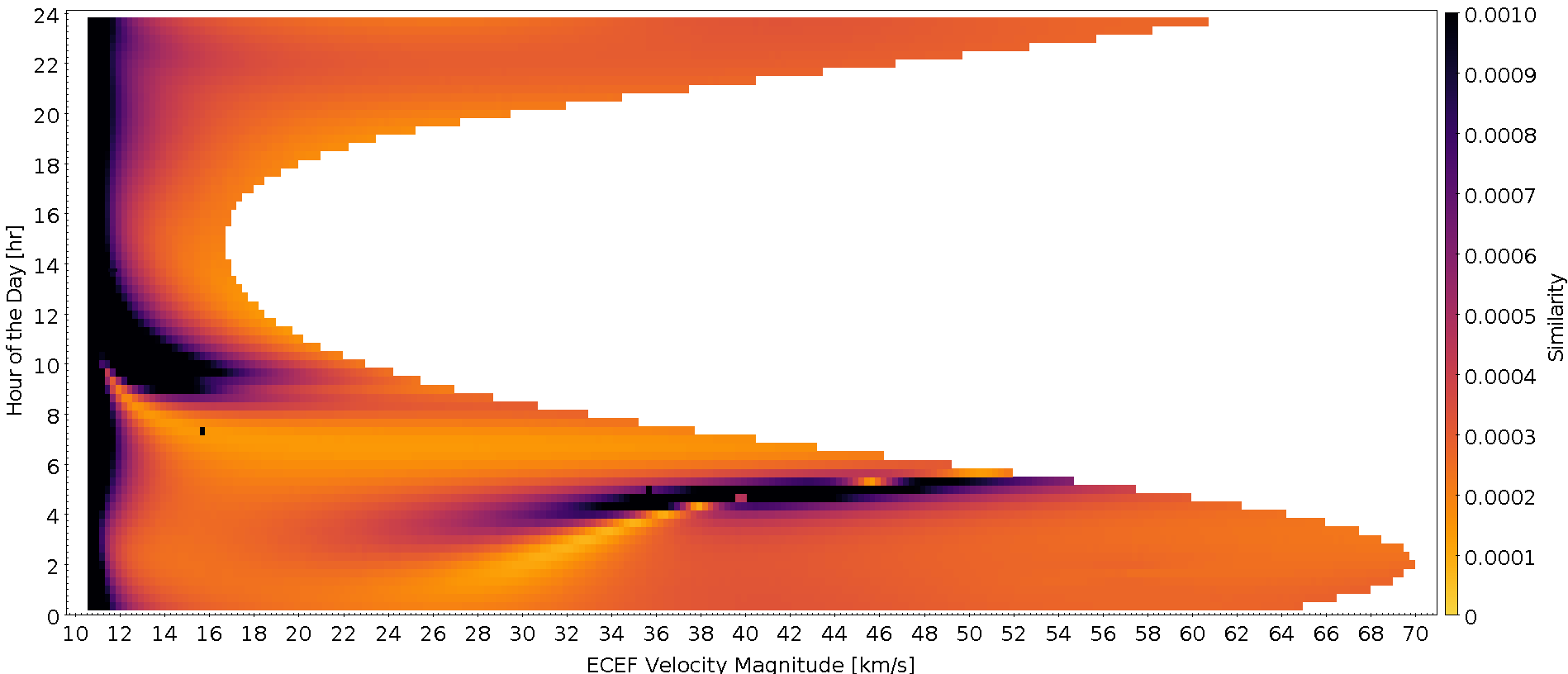}
    \caption{Orbital similarity between Ceplecha’s analytical method (C-87) and the new numerical method described in this paper (JS-19) according to the Southworth and Hawkins similarity criterion \citep{southworth_statistics_1963}. The darker the shade, the more difference there is between the simulated orbits. Only the heliocentric orbits are shown; all hyperbolic and geocentric orbits are discarded. The sinusoidal-like shape is due to the orbital velocity of the Earth around the Sun. The two distinctly darker areas at lower velocities represent strong perturbations that are not considered in the C-87 model. }
    \label{fig:similarity}
\end{figure}

A further assessment of similarity between C-87 and JS-19 can be made beyond the single observed Hayabusa re-entry using a variety of simulated re-entry trajectories. We can generate simulated trajectories using the Earth fixed re-entry radiant unit vector of the Hayabusa satellite as the trajectory backbone. This is then varied by artificially altering the velocity magnitude and the time of re-entry.  By modifying the re-entry time, we are effectively adjusting the longitude of the re-entry in an inertial frame due to the Earth’s diurnal rotation. We vary the re-entry time through an entire day in $20\,minute$ increments, given in UTC time. At each of these discrete time increments, the re-entry velocity magnitude is also varied to cover all possible heliocentric orbits conservatively, i.e. from $10\,km/s$ up to $80\,km/s$ in $250\,m/s$ increments; any resulting hyperbolic orbits are dismissed. On each of the 2,088 simulated trajectories within this dataset, the orbit is computed once using C-87 and once using JS-19. The similarities of the determined heliocentric orbits are shown in Fig.~\ref{fig:similarity}. 

The general shape of Fig.~\ref{fig:similarity} is due to the Earth’s velocity around the Sun. At about 15:00\,UTC on June 13th, 2010, the Earth’s velocity acts in the same direction as the simulated Hayabusa re-entry, therefore reducing the velocity needed to obtain a hyperbolic orbit relative to Earth. Conversely, around 03:00\,UTC, the simulated velocity relative to Earth must be much higher to obtain a hyperbolic orbit as the Earth’s velocity opposes the simulated Hayabusa re-entry velocity. Additionally, the minimum Earth centred velocity needed to obtain a heliocentric orbit is the Earth’s escape velocity, regardless of the Earth’s orientation around the Sun. 

\begin{figure}
    \centering
    \includegraphics[width=\textwidth]{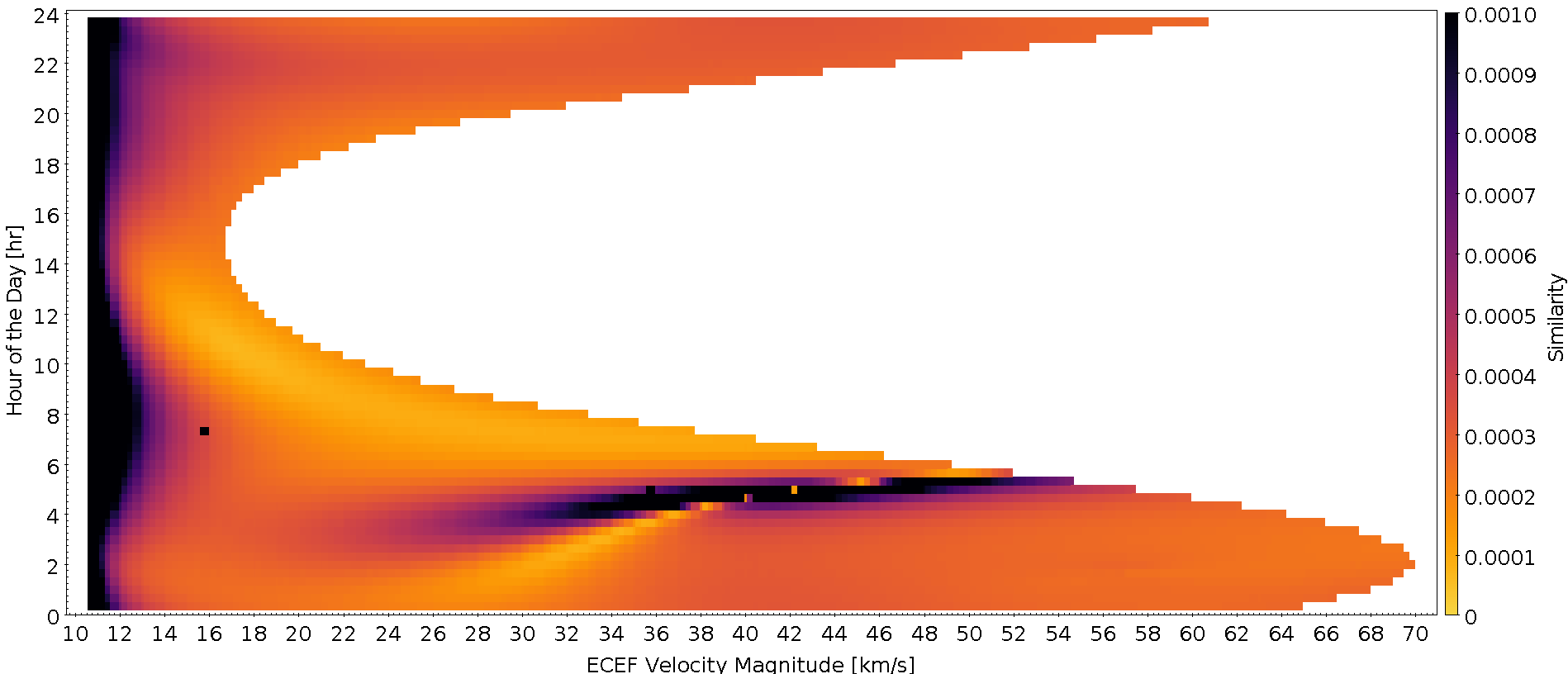}
    \caption{Orbital similarity \citep{southworth_statistics_1963} between Ceplecha’s analytical method (C-87) and the new numerical method described in this paper (JS-19) having removed the Moon’s perturbation influence from the latter. Note the removal of the lunar effect between 08:00-12:00\,UTC from Fig.~\ref{fig:similarity}.}
    \label{fig:similarity_no_moon}
\end{figure}

Interestingly, certain regions of orbital dissimilarity can be identified by excluding particular perturbations from JS-19. For example, by removing the Moon’s gravitational perturbing influence, the orbit produced by the numerical algorithm becomes more like the orbit produced by C-87 in the area between 08:00-12:00\,UTC (Fig.~\ref{fig:similarity_no_moon}).

\begin{figure}
    \centering
    \includegraphics[width=\textwidth]{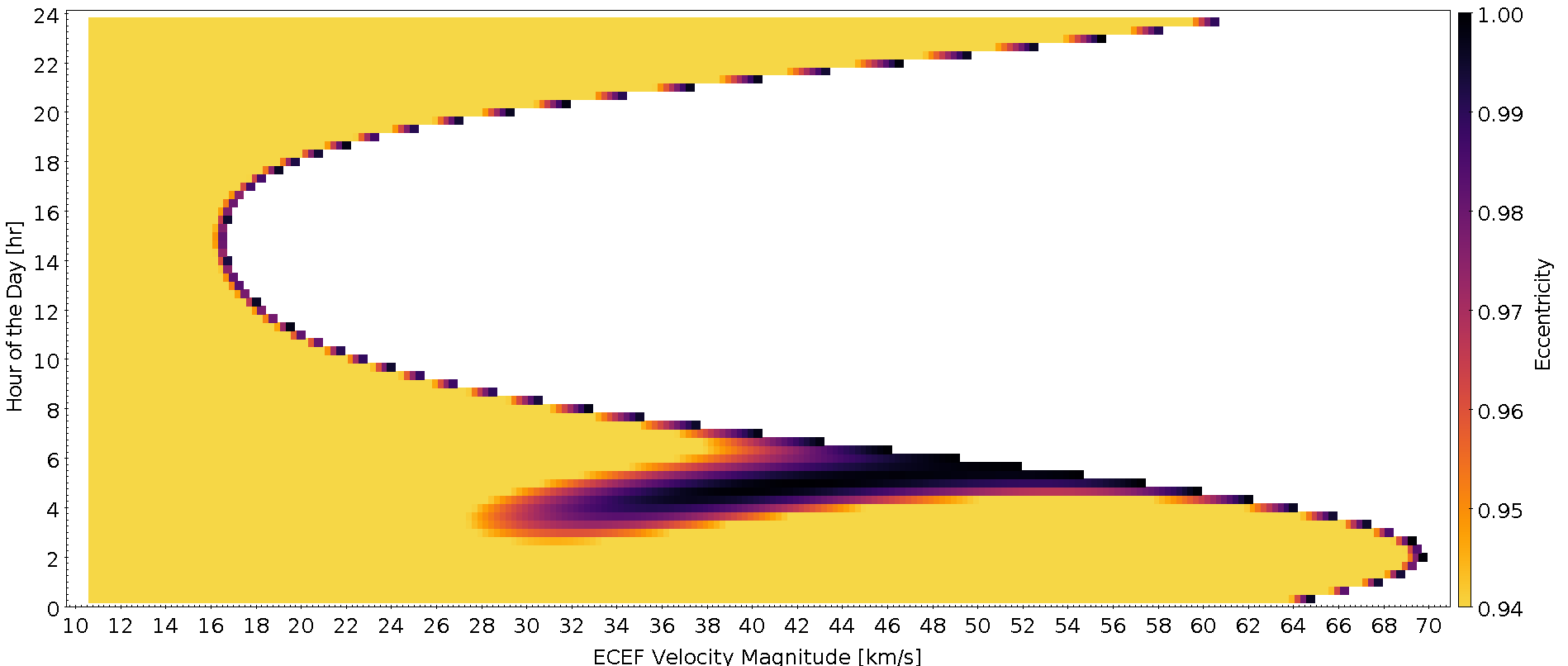}
    \caption{The region of high eccentricity for the simulated dataset of re-entry trajectories.}
    \label{fig:similarity_eccentricity}
\end{figure}

Other regions of orbital dissimilarity in Fig.~\ref{fig:similarity} and Fig.~\ref{fig:similarity_no_moon} can also be identified. The darker region at lower re-entry velocities is due to the resulting orbit being close to that of the Earth’s orbit, and therefore experiencing a greater time for the Earth/Moon perturbations to influence the orbit off its Keplerian path. Additionally, the roughly horizontal region at higher re-entry velocities, around 05:00\,UTC, corresponds to an area of high orbital eccentricity (Fig.~\ref{fig:similarity_eccentricity}). As \citet{jopek_remarks_1993} describes, the values of the similarity criterion \citep{southworth_statistics_1963} “strongly depend on the orbit eccentricity when $e > 0.9$”, therefore accounting for this region of apparent dissimilarity.

Also note, the isolated dots in Fig.~\ref{fig:similarity} and Fig.~\ref{fig:similarity_no_moon} are single orbital cases where the inclination is so close to zero that the calculated longitude of ascending node, $\Omega$, in one orbital estimation is the longitude of descending node, \rotatebox[origin=c]{180}{$\Omega$}, in the other. This results in a misdiagnosis of orbital similarity. 

To give the reader an idea of these orbital differences, if the velocity magnitude uncertainty was the only acting source of orbit error, then $1\,m/s$, $10\,m/s$, $100\,m/s$ uncertainties on initial velocity magnitude would correspond to orbital similarities of 0.0002, 0.002, and 0.02, respectively. These correlations are specific to the geometry of the Hayabusa trajectory and may vary for different events, but serve well as a rough similarity conversion for Figures~\ref{fig:similarity} and \ref{fig:similarity_no_moon}. That said, the similarity range in Figures~\ref{fig:similarity} and \ref{fig:similarity_no_moon} are capped at 0.001 to highlight subtleties, however some orbit comparisons, especially at the low velocity end, did show similarities on the order of 0.02, or roughly $100\,m/s$ variation in initial velocity magnitude; a significant difference in orbital terms.

This analysis highlights differences in orbit determination methods due to re-entry timing and velocity magnitude, and even further differences may be caused by variations in re-entry height, latitude, azimuth and zenith angles.

So, without the ability to include perturbations, C-87 cannot properly account for the complexities inherent in the estimation of pre-Earth orbits. Any discrepancy from the meteoroid’s ‘true’ orbit will be magnified when a probabilistic method, such as \citet{bottke_effect_2002}, is used to determine its orbital origins, therefore making it significantly harder to link meteoroids to their rightful parent bodies or source regions.

\section{Conclusions}
Ceplecha’s analytical method of orbit determination (C-87) is computationally easy, and historically the most widely used technique in determining the originating orbits of meteoroids. However, it does not allow for perturbations in orbit calculations such as third bodies (including the Moon) or Earth flattening effects. A numerical approach is able to incorporate such perturbations. With increasing computational power, such an approach is preferable. 

A new numerical method (JS-19) is presented in this study. To compare the results of this new orbital determination technique to the typical analytical method (C-87) and the numerical approach provided in the Meteor Toolkit package (MT-15), the re-entry observations of JAXA’s Hayabusa, with its known heliocentric orbit as a ‘ground truth’, was invaluable. As observations were made of both the spacecraft and the capsule re-entry separately, these data provide two excellent test cases with which models could be compared to heliocentric telemetry. The spacecraft was first observed at $\sim100\,km$ altitude while the capsule was not observed until $\sim65\,km$ altitude. The low observation altitude of the capsule tests the capability of models to incorporate atmospheric influences. In both cases, JS-19 determined the most similar orbit to JAXA’s recorded orbit than either C-87 or MT-15. This was especially evident in the second case due to the greater atmospheric influence that the capsule experienced before initial sighting. Further investigation of the atmospheric influence shows the need for atmospheric consideration in meteoroid orbit determination below $\sim90\,km$ altitude. This is therefore highly relevant for many meteorite dropping events which may not be initially observed above this height by fireball networks tuned to brighter events. We also stressed that C-87 alone does not account for atmospheric drag effects, requiring a pre-atmospheric initial velocity to be determined prior to its use. The calculation of this initial velocity by the majority of current fireball networks that use C-87 is unclear and may need to be revised.

We made a detailed assessment on the accuracy and precision of orbital calculations. The numerical methods are shown to produce more realistic precision and deliver superior accuracy in estimating the Hayabusa spacecraft’s pre-Earth orbit from re-entry observations than the analytical method, verifying such claims of previous authors \citep{clark_numerical_2011, jenniskens_cams:_2011}.

The resulting orbital element precision is primarily determined by the size of the initial velocity magnitude error, as all other foreseeable uncertainties combined correspond to orbital errors at least an order of magnitude smaller than the initial speed uncertainty, as discussed in Section~\ref{sssec:precision}. While the precision of the orbit determination methods were comparable, JS-19 demonstrated greater accuracy due to its complete detailed representation of Earth’s gravity and its inclusion of perturbations, as shown in Section~\ref{sssec:accuracy}. 

By generating a great variety of simulated re-entry trajectories, we were able to explore the effect of different perturbations by comparing orbits calculated by both C-87 and JS-19. Simulated trajectories with low entry velocities or which pass close to the Moon show the most drastic orbital divergences. This demonstrates the vital need for perturbation inclusion within the orbit determination method. The limitations of C-87 should be considered and discussed if used for meteoroid orbit determination. Previously determined orbits, especially those in regions of significant orbital divergence (as discussed in Section~\ref{sssec:accuracy}) should be re-analysed to avoid inaccurate orbital histories.

The Hayabusa case used in this work has provided a unique opportunity to compare orbit determination techniques. Although this case assesses only a heliocentric orbit, it must be noted that JS-19 can compute an observed meteoroid’s orbit regardless of whether it originated around the Earth (geocentric), around the Sun (heliocentric), or from outside the solar system (hyperbolic), proving itself to be a more robust and real world approach than its analytical counterpart.

\section{Acknowledgements}
We would like to specially thank Dr. Makato Yoshikawa and the Hayabusa engineering team for providing the telemetry data for the spacecraft that made this study possible. Thank you to M. Gritsevich, J. Borovicka, and an anonymous reviewer for your valuable feedback which helped improve the quality of this manuscript. We also thank the Australian Research Council for support under their Australian Discovery Project scheme. This research was supported by an Australian Government Research Training Program (RTP) Scholarship.

\section{References}
    \bibliography{research}{}
    \bibliographystyle{abbrvnat}

\pagebreak
\section{Appendix: Initial Conditions}
The initial conditions for our comparative analysis are determined using the tabulated values given in \citet{borovicka_photographic_2011}, but are collated in the table below. Note the quoted velocities are relative to the ground (ECEF frame).

\begin{table}[h!] 
    \centering
    \begin{threeparttable}
    \begin{tabular}{ |l||c|c| }
        \hline
                                            & \textbf{Spacecraft}   & \textbf{Capsule}      \\
        \hline
        Epoch Time                          & 2010-06-13T13:51.56.6 & 2010-06-13T13:52.16.0 \\
        Latitude [deg]                      & -29.0243              & -29.6545              \\
        Longitude [deg]                     & 131.1056              & 133.0768              \\
        Height [km]                         & 99.880                & 64.710                \\
        Initial velocity [m/s]              & 11725.1               & 11330.5               \\
        Radiant azimuth [deg]               & 290.5220              & 289.2733              \\
        Radiant elevation [deg]             & 10.0173               & 8.7955                \\
        Mass [kg]                           & 415                   & 20                    \\
        Cross-sectional Area [m2]\tnote{a}  & 2.15                  & 0.126                 \\
        Corresponding radius [m]            & 0.827                 & 0.2                   \\
        Infinite velocity [m/s]\tnote{b}    & 11678.84              & 11939.04              \\
        \hline
    \end{tabular}
    \begin{tablenotes}\begin{footnotesize}
        \item[a] Using a drag coefficient of 2.
        \item[b] Using methods described in the appendix of \citet{pecina_new_1983}.
    \end{footnotesize}\end{tablenotes}
    \end{threeparttable}
    \label{table:orbits}
\end{table}

\end{document}